\documentclass{article}

\usepackage{arxiv}

\usepackage[utf8]{inputenc} 
\usepackage[T1]{fontenc}    
\usepackage{hyperref}       
\usepackage{url}            
\usepackage{booktabs}       
\usepackage{amsfonts}       
\usepackage{nicefrac}       
\usepackage{microtype}      
\usepackage{lipsum}
\usepackage{graphicx}
\usepackage{framed,multirow}
\usepackage{amssymb}
\usepackage{latexsym}
\title{DeepCervix: A Deep Learning-based Framework for the Classification of Cervical Cells Using Hybrid Deep Feature Fusion Techniques}

\author{
  Md Mamunur Rahaman\\
  \\
  Microscopic Image and Medical Image Analysis Group, 
MBIE College\\ Northeastern University\\ Shenyang 110169, China\\
  \texttt{mamunrobi35@gmail.com} \\
   \And
 Chen Li \\
 Microscopic Image and Medical Image Analysis Group, 
MBIE College\\ Northeastern University\\ Shenyang 110169, China\\ \texttt{lichen201096@hotmail.com}\\
\And
Yudong Yao \\
Department of Electrical and Computer Engineering\\ 
Stevens Institute of Technology\\ Hoboken, NJ 07030, USA\\
\And
Frank Kulwa\\
Microscopic Image and Medical Image Analysis Group\\ 
MBIE College\\ Northeastern University\\ Shenyang 110169, China\\
\And
Xiangchen Wu\\
Suzhou Ruiguan Technology Company Ltd.\\ Suzhou 215000, China\\
\And
Xiaoyan Li \\
Cancer Hospital of China Medical University\\ Liaoning Hospital and Institute\\ 
Shenyang 110042, China\\
\And
Qian Wang \\
Cancer Hospital of China Medical University\\ Liaoning Hospital and Institute\\ 
Shenyang 110042, China\\
}

\begin{document}
\maketitle

\begin{abstract}
Cervical cancer, one of the most common fatal cancers among women, can be prevented by regular screening to detect any precancerous lesions at early stages and treat them. Pap smear test is a widely performed screening technique for early detection of cervical cancer, whereas this manual screening method suffers from high false-positive results because of human errors. To improve the manual screening practice, machine learning (ML) and deep learning (DL) based computer-aided diagnostic (CAD) systems have been investigated widely to classify cervical pap cells. Most of the existing researches require pre-segmented images to obtain good classification results, whereas accurate cervical cell segmentation is challenging because of cell clustering. Some studies rely on handcrafted features, which cannot guarantee the classification stage's optimality. Moreover, DL provides poor performance for a multiclass classification task when there is an uneven distribution of data, which is prevalent in the cervical cell dataset. This investigation has addressed those limitations by proposing DeepCervix, a hybrid deep feature fusion (HDFF) technique based on DL to classify the cervical cells accurately. Our proposed method uses various DL models to capture more potential information to enhance classification performance. Our proposed HDFF method is tested on the publicly available SIPAKMED dataset and compared the performance with base DL models and the LF method. For the SIPAKMED dataset, we have obtained the state of the art classification accuracy of $99.85\%$, $99.38\%$, and $99.14\%$ for 2-class, 3-class, and 5-class classification. Moreover, our method is tested on the Herlev dataset and achieves an accuracy of $98.32\%$ for binary class and $90.32\%$  for 7-class classification.
\end{abstract}


\section{Introduction}
\section{Introduction}
\label{sec:introduction}
Cervical cancer, found in woman's cervix, is the fourth most prevalent cancer among women~\cite{rahaman2020survey}. According to the World Health Organization (WHO), approximately 570 000 women are diagnosed with cervical cancer globally, and about 311 000 women have lost their lives due to this fatal disease in 2018 alone~\cite{world2019guidelines}. More than 80\% of the cervical cancer cases and 85\% of deaths occur in poor and developing nations because of the absence of screening and treatment facilities~\cite{ferlay2018global}. Improper menstrual hygiene, pregnancy at an early age, smoking and use of oral preventatives are the leading risk factors that lead to the infection with human papillomavirus (HPV)~\cite{vsarenac2019cervical}. Research has revealed that long term infection with HPV is the main reason for cervical cancer. However, Cervical cancer is the most treatable form of cancer if it is detected early and treated adequately~\cite{saslow2012american}.  

Routine screening of women over 30 years old plays a vital role to prevent cervical cancer effectively by allowing the early detection and treatment~\cite{world2014screening}. The most popular screening technique to detect the cervical malignancy is cervical cytopathology (pap smear test or liquid-based cytology) due to its cost-effectiveness~\cite{saslow2012american, davey2006effect}. In this technique, cells are collected from the squamocolumnar terminal of the cervix and the malignancy is checked under the light microscope by expert cytologists~\cite{papanicolaou1973new, papanicolaou1941diagnostic}. It usually demands 5-10 minutes to analyze a single slide based on the different orientation and overlapping of the cells~\cite{elsheikh2013american}. Moreover, manual screening method is difficult, tedious, time-consuming, expensive and subject to errors because each slide contains around three million cells with different orientation and overlapping, which leads to developing an automated computerized system that can analyze the pap cell effectively and efficiently~\cite{gencctav2012unsupervised,lozano2007comparison}.

With the possibility to train data at the end of 1990s, there has been extensive research for the development of computer-aided diagnostic (CAD) system to help doctors to track cervical cancer~\cite{litjens2017survey}. The traditional CAD system consists of three steps: cell segmentation (cytoplasm, nuclei), feature extraction and classification. In this system, firstly, filtering based preprocessing work is performed to enhance image quality. Then, cell nuclei are extracted using k-means~\cite{krishna1999genetic}, clustering~\cite{kanungo2002efficient} or super-pixel~\cite{lee2016segmentation} methods. After, the post processing task is performed to correct the segmented nucleus. After that, handcrafted features~\cite{jantzen2005pap, marinakis2008particle, marinakis2009pap}, such as Morphological features, color metric features and texture features are extracted from the segmented nucleus. Next, the feature selection technique is applied to find the most discriminant features, and finally, a classifier is designed to classify the cell~\cite{win2018computer}.

\begin{figure*}[t!]
\centering
\includegraphics[scale=.7]{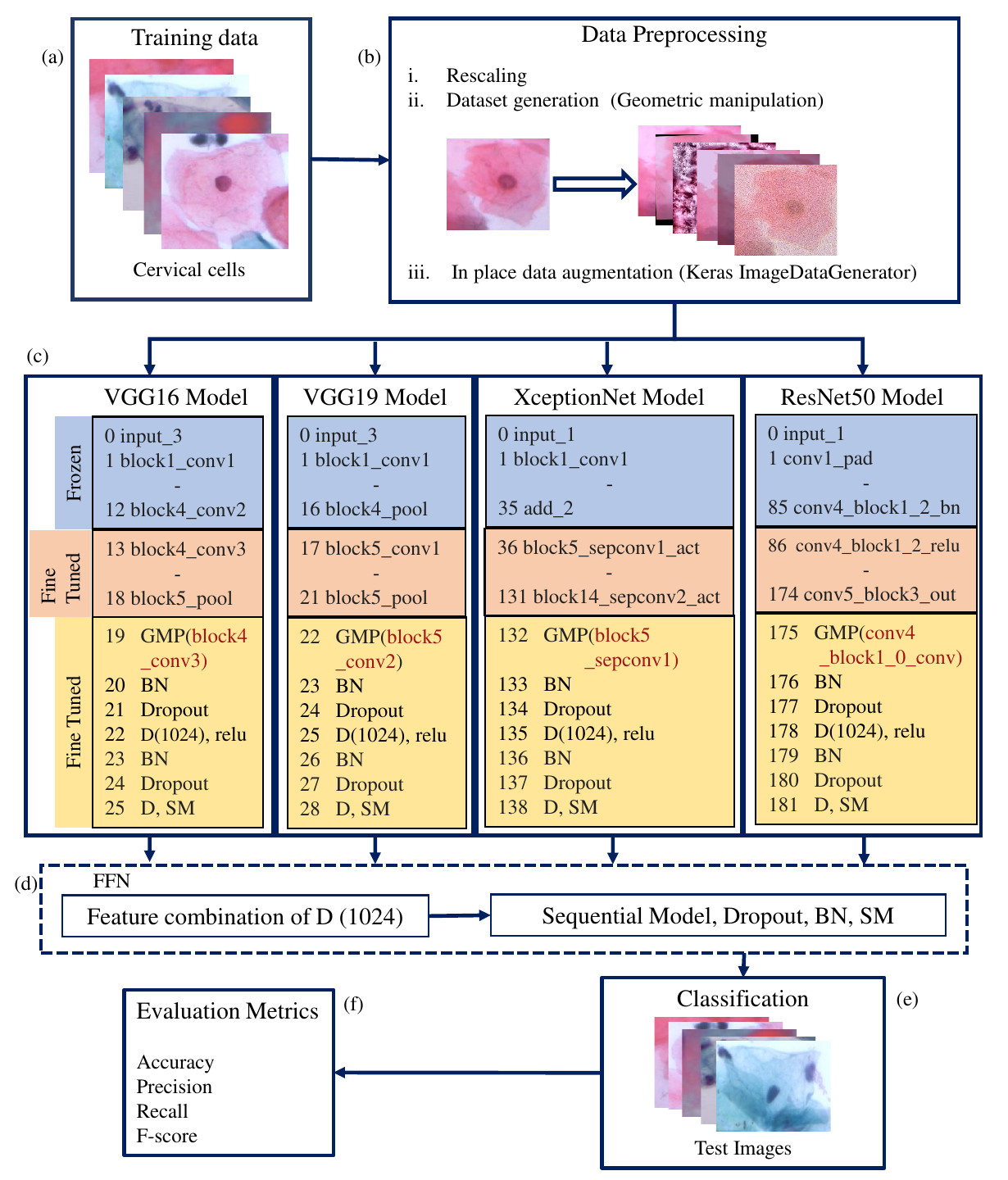}
\caption{Workflow diagram of the proposed DeepCervix network. (Global Max Pooling (GMP), Batch Normalization (BN), Dense Layer (D), SoftMax (SM))}
\label{workflow}
\end{figure*}

The above-described method requires many steps to process the data and extracted handcrafted features cannot ensure superior classification performance, which also highlights the incompetence of automatic learning. In order to obtain an enhanced CAD system, deep learning (DL) based feature extraction methods have a significant advantage over other machine learning (ML) algorithms. DL based algorithm is achieving the state-of-the-art results on challenging computer vision tasks~\cite{goodfellow2016deep, russakovsky2015imagenet}.  One compromise with DL is that it demands a considerable amount of data to obtain a good result compared with ML techniques, which is challenging to obtain in the medical domain~\cite{landau2019artificial}. Moreover, DL also provides poor performance when there is an uneven distribution of the sample data in a multiclass classification problem, which is very prevalent in the medical domain. Therefore, the CAD technique for the analysis of pap cells requires further research and development.

In this study, we have introduced DeepCervix, which is a DL based framework to accurately classify the cervical cytopathology cell based on hybrid deep feature fusion (HDFF) techniques. In our proposed framework, we have used pre-trained DL models that are trained on ImageNet datasets ($>$1 million images) and then fine-tuned it on the cervical cell dataset, which resolves the requirement of plenty of datasets and challenges associated with multiclass classification with uneven data distributions. Moreover, deep feature fusion (DFF) from various DL models is capable of capturing more potential information, which improves the classification performance. Our proposed method is tested on SIPAKMED dataset, consisting of single-cell cervical cytopathology images. For SIPAKMED dataset, we have achieved the highest classification accuracy of 99.85\%, 98.38\% and 99.14\% for 2-class, 3-class and 5-class classification problems, respectively. Moreover, we have also tested our method on Herlev dataset and reached an exactitude of 98.91\% for binary classification and 90.32\% for 7-class distribution problem. The workflow of the suggested HDFF method is presented in Fig.~\ref{workflow}. From the workflow diagram, we can see that: 
\begin{itemize}
\item As shown in Fig.~\ref{workflow}, the cervical pap smear images are first retrieved from accessible databases (e.g., SIPAKMED, Herlev) and considered as training samples.\\
\item In the preprocessing step, two stages of data augmentation task are implemented; first is to use some geometric manipulation, such as affine transformations, adding noises (Gaussian, Laplace), canny filter, edge detection, colour filter, change of brightness and contrast to increase the training samples. Second is to use the in-place data augmentation technique utilizing the Keras “ImageDataGenerator” API, where the images are reconstructed randomly during the training time.\\
\item After the preprocessing step, the images are supplied to four DL models, VGG16, VGG19, XceptionNet and ResNet50. From Fig.~\ref{workflow}-(c), it is seen that for VGG16 model, we have fine-tuned the last convolutional block, from layer-13 to layer-18 along with the top-level classifier.\\
\item In the feature fusion network (FFN) stage, first, we extract the features from the last layer before the SM layer of the DL models to create the feature arrays with 1024 features from each model. Then, the feature arrays are fed into the sequential model connecting with dense layer with BN and dropout layer in between, to perform the classification.\\
\item In this step, unseen test images are provided to perform the classification.\\
\item Finally, we have assessed the performance of the proposed model by calculating the precision, recall, $F1$ score and accuracy.
\end{itemize}

The main contributions of this paper are as follows: 
(1) To the best of our knowledge, this is the first study to classify cervical cytopathology cell using HDFF techniques. 
(2) Two different stages of data augmentation techniques are presented in this study.
(3) Four types of CNN’s with enhanced structure, VGG16, VGG19, XceptionNet and ResNet50 are introduced to extract the complementary features from various depths of the networks. 
(4) An improved FFN is included to integrate the features adaptively by combining dense layer with SM, BN and dropout layer in between.
(5) Our proposed method achieves the highest classification accuracy on the SIPAKMED dataset, which shows the potential of improved cervical cancer diagnostic systems.

The remainder of this paper is organized as follows: Sec.~\ref{sec:literature} presents relevant studies of DL for the analysis of cervical cytopathology images and relevant feature fusion studies in computer vision tasks. Sec.~\ref{sec:method} investigates data pre-processing techniques that we have utilized in our experiment and our proposed methods. Sec.~\ref{sec:experiment} explains the experimental dataset, data settings, experimental setup, evaluation method, and experimental results and analysis. Sec.~\ref{discu} discusses our proposed method with some examples of misclassified images. Finally, Sec.~\ref{sec:conclu} concludes this paper by pointing out some limitations of our method.

\section{Literature Review}\label{sec:literature}
An overview of relevant DL approaches that are employed to analyze the cervical cells and feature fusion techniques in imaging modalities are compiled in this section.

\subsection{Relevant investigations of DL for the analysis of cervical cytopathology images}
Various DL and ML-based techniques have been applied to classify the cervical cells. For instance,~\cite{sukumar2015computer} utilizes the histogram features, texture features, grey level features and local binary pattern features. Then, the features are supplied into a hybrid classifier system combining with SVM and adaptive neuro-fuzzy interface system to analyze the cervical cells into normal and abnormal. A hybrid ensemble technique is introduced by combining 15 different machine learning algorithms, such as random forest, bagging, rotation forest and $J48$ graft to classify the cervical cells~\cite{sarwar2015hybrid}. They observe that a hybrid ensemble technique performs better than an individual algorithm.

A deep CNN (base AlexNet) based feature extraction method is applied in~\cite{bora2016pap}, followed by an unsupervised feature selection task. Later, feature vectors are supplied into the least-square version of the support vector machine (LSSVM) and SoftMax regression to classify the cervical cells. \cite{hyeon2017automating} designs a model to extract the features using VGG16 from cervical cells and fed the features into ML classifiers, support vector machine (SVM), random forest and AdaBoost. They discern that SVM functions better than other ML classifiers. A pre-trained AlexNet architecture is employed to extract the characteristics of cervical cells and apply those features to classify them using SVM~\cite{taha2017classification}. A CNN based classification approach is explained in~\cite{wieslander2017deep} to classify the cervical cells applying VGG16 and ResNet architecture and explore that ResNet50 is more suitable than VGG16 based on the performance. A deep transfer learning-based classification approach is presented in~\cite{zhang2017deeppap} to classify the cervical cells into healthy and abnormal with prior data augmentation and patch extraction work. \cite{gautam2018considerations} applies deep transfer learning technique based on AlexNet to detect, segment and classify the cervical cells and demonstrates that segmentation is not necessary for classification. AlexNet, GoogleNet, ResNet and DenseNet based pre-trained and fine-tuned CNN architecture is employed to classify the cervical cells in~\cite{lin2019fine}, where segmentation of cytoplasm and nucleus are prerequired for this method.

Similarly, In~\cite{allehaibi2019segmentation}, VGG-like network consists of seven layers uses pre-segmented cervical cells to perform the classification task. A comparative study is performed based on five DL models, ResNet101, Densenet161, Alexnet, VGG19 and SqueezeNet to check their classification performance on the cervical dataset, where DenseNet161 provides the maximum accuracy~\cite{promworn2019comparisons}. Moreover, \cite{nguyen2019biomedical} coupled the features of pre-trained Inception-V3, ResNet152 and InceptionResNetV2 to analyze biomedical images. In addition, a detailed study about relevant work, it is recommended to go through our survey paper about cervical cytopathology image analysis using DL~\cite{rahaman2020survey}.

It is perceived from the reference review that most of the authors have conducted a binary classification task, whereas, in practice, multiclass classification is more important. Moreover, the transferred model often unable to acknowledge the characteristics of medical images, and traditional features can not guarantee the optimality of the system. Therefore, this paper investigate methods to address those issues.
\subsection{Relevant investigation of feature fusion in computer vision tasks}
A hybrid fusion approach, combining early and late fusion is presented in~\cite{benzebouchi2019multi} for the diagnosis of glaucoma. Handcrafted features such as Gray level co-occurrence matrix, central and Hu moments are consolidated with deep features. Later, the feature vectors are supplied to SVM and CNN based classifier. A satellite remote sensing scene classification method based on multi-structure deep feature fusion is presented in~\cite{xue2020remote}. CaffeNet, VGG-VD16 and GoogLeNet are applied to extract the features and fuse those features through the fusion network to do the classification. \cite{wang2019breast} develops a CAD method to detect breast cancer by employing feature fusion with CNN. They have combined the deep features, morphological features, texture features, density features and fuse those features through extreme machine learning classifier to classify the breast masses into benign and malignant. In our previous study~\cite{xue2020application}, we have classified cervical histopathology images using weighted voting based ensemble learning techniques. In~\cite{kumar2016ensemble}, an ensemble of different CNN structure, is obtained to classify medical images. The proposed ensemble method proves better predictive capability by combining the results of different classifiers. \cite{amin2020integrated} practices the pre-trained AlexNet and VGG16 to extract the features from segmented skin lesions and classify them into benign and malignant.

\section{Method}\label{sec:method}
\subsection{Data Preprocessing}
\subsubsection{Rescaling}
The cervical cytopathology cell images (SIPAKMED dataset) that we have employed to check the performance of our proposed method are in BMP format with dimensions ranging from ($71 \times 59$) to ($490 \times 474$) pixels. Therefore, we have rescaled the object size to ($224 \times 224$) pixels for all the four CNN networks. In this respect, we have utilized the Keras ``preprocess-input" function, which transforms input images according to the model requirement.

\subsubsection{Dataset generation}
Various geometric transformations and image processing functions are discussed in this subsection that we have used in our experiment. The data augmentation task is performed using machine learning ``imgaug" library, fourth version, which supports various augmentation techniques. The newly formed images saved along with the training images and increase the training data size by a factor of six, which is used to obtain better results.

\begin{itemize}
\item Affine Transformations (ATs): ATs are geometric manipulations that move a pixel from a coordinate position of $(a, b)$ to a new position of $(a^\prime, b^\prime)$. A pair of transformations specify the movement,
\begin{equation}
a^\prime = T_a(a,b),       b^\prime = T_b(a,b)
\end{equation}
It combines linear transformations and translations. In our experiment, we have performed rotation, scaling, translation, shearing and horizontal and vertical flip operations of an image. For a batch of training images, one of these transformations is randomly arranged. 
\item Contrast limited adaptive histogram equalization (CLAHE):
As we know, histogram equalization (HE) enhances the contrast of images, which may lead to too bright or dark regions. Whereas, CLAHE performs histogram equalization by dividing images into small blocks, where each block performs HE. As a result, it prevents the over-amplification of noise and contrast in an image. CLAHE, all channel CLAHE and gamma contrast are employed in our experiment. One of the CLAHE augmenters is randomly chosen from a batch of training samples.
\item Edge detection: ``EdgeDetect'' and ``DirectedEdgeDetect" functions are used from imgaug API that transforms the input images into edge images, where edges are detected from random angles and mark non-edge region as black and edge region as white. 
\item Canny filter: Canny edge detection augmenters are also utilized, where the input images are preprocessed using Sobel filter.
\item Photometric transformations (Pms): PMs are accomplished by shuffling all the colour channels, turning images into grayscale, changing hue and saturation value, adding hue and saturation and quantizing images up to 16 colours.
\item Contrast adaptation (CA): CA is performed by modifying the contrast and brightness of an image. 
\end{itemize}

\subsubsection{In place data augmentation}
In order to enhance model performance, Keras ``ImageDataGenerator" API is applied~\cite{rahaman2020identification}. The images are transformed randomly during the training time. As a result, the network examines unlike samples in each epoch, which extend the model generalizability. In this process, we have set the featurewise center as false, rotation range is set to 5 degrees and fill mode is nearest. Then, we have fixed horizontal and vertical flips to true, brightness range from 50\% to 130\% and kept the channel shift range true.

\subsection{Basic methods}
\subsubsection{Deep learning}

Lately, DL, one type of ML algorithms, is the most commonly designed and successful type of ml algorithm to analyze the medical images. Convolutional neural network (CNN) is the most prevalent deep learning architecture. Research has confirmed that CNNs are robust to image noise and invariant to translation, rotation and size, which increase the object's analyzing ability~\cite{rolnick2017deep, dieleman2015rotation}. The CNN architecture is composed of convolution, pooling and fully connected layers. The main building block of CNN structure is convolution layer, which extracts the low- and high-level features of an image as the layer gets deeper~\cite{khan2020survey}. The pooling layer after the convolution layer reduces the size of the convoluted features by extracting the maximum or average value through max-pooling or average pooling operation. A fully connected layer (FCL) connects every neuron of each layer to another layer to classify the image, followed by the principle of multilayer perceptron~\cite{suarez2018evaluation}. In this study, we have utilized VGG-16, VGG-19, ResNet-50 and XceptionNet as CNN architecture.

\begin{enumerate}

\item VGGNet: The VGGNet came with the idea of a deeper network with smaller filter. The model can have 16 to 19 layers with fixed input size of $224 \times 224 \times 3$. The convolution filter size is ($3 \times 3$) with a stride of 1 pixel. A linear transformation of input is also performed by ($1 \times 1$) convolution filter with ReLU activation function. A total of five max-pooling operations is performed with window size ($2\times2$), followed by three FCL. The significant discovery of the VGGNet is the small receptive field ($3\times3$), which enables to have more weight layers, consequently, to improve the performance~\cite{simonyan2014very}.

\item ResNet: \cite{he2016deep} observes that with the increase of network depth the network performance improves at a certain level and then degrades rapidly. Therefore, it introduced skip connections to increase the performance with network depth. Thus, it is possible to have 1000 weight layer in ResNet. For a X feature input of a convolution layer with F(x) as a residual function, the input of the first layer (x) is copied to the output layer, 
\begin{equation}
H(x) = F(x) + x, or, F(x) = H(x) - x
\end{equation}
The structure of the residual learning block is shown in Fig.~\ref{learning}.

\item XceptionNet: The extended version of Inception model is XceptionNet, which is based on depth wise separable convolutions, followed by pointwise convolution. The model is lighter with few number of connections and provides better results on ImageNet classification then InceptionV3, ResNet and VGGNet~\cite{chollet2017xception}.

\begin{figure}
\centering
\includegraphics[scale=.8]{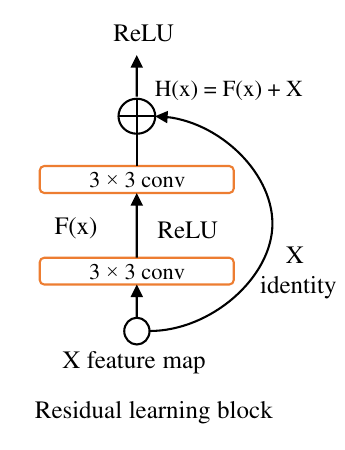}
\caption{The structure of residual learning block of Resnet.}
\label{learning}
\end{figure}
\end{enumerate}

\subsubsection{Transfer learning}
\begin{figure*}[th!]
\centering
\includegraphics[scale=.75]{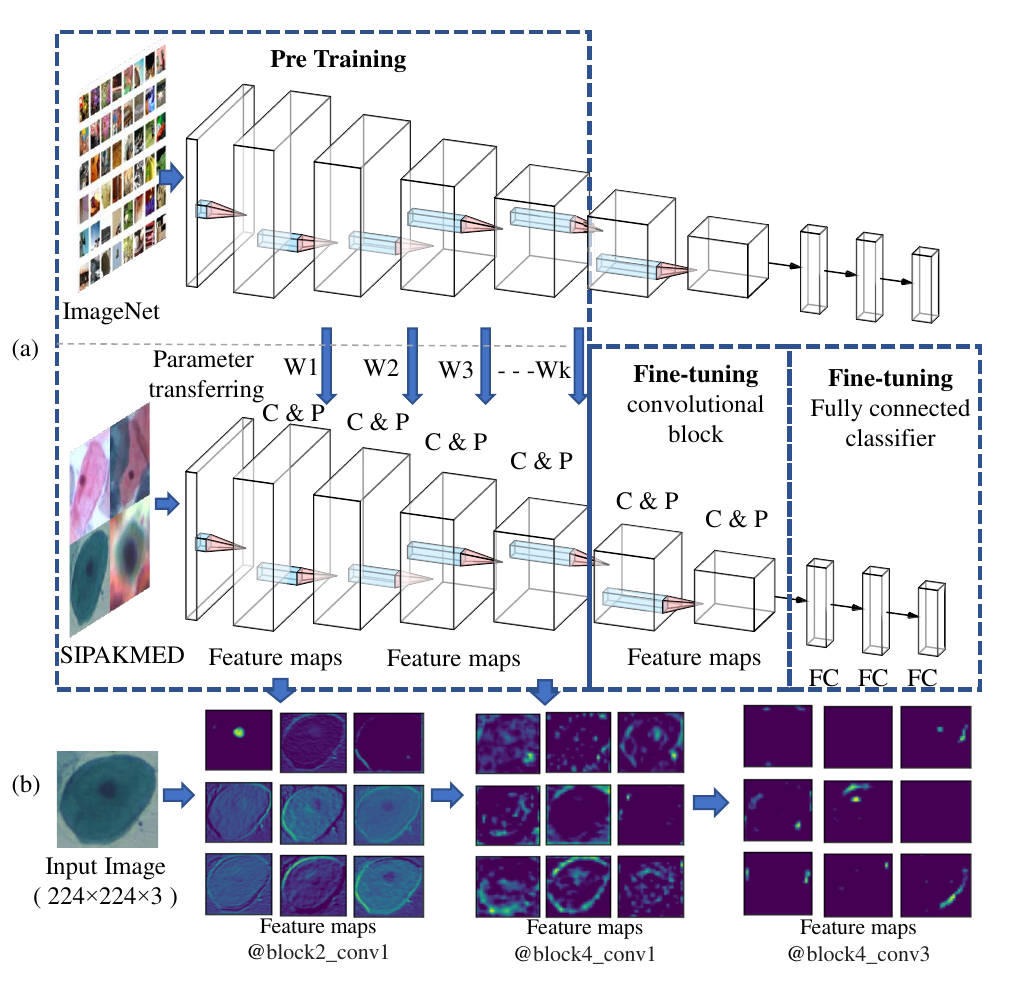}
\caption{(a) Visualization of TL process, where parameters are transferred from another CNN and fine-tuned on cervical cancer cell dataset, (b) Visualization of the feature maps of three different convolutional layers of VGG16.}
\label{transfer}
\end{figure*}
To train a CNN from scratch demands a considerable amount of data with high computing power, which also costs longer training time. In medical domain, image datasets are usually in the order of $10^2-10^4$, since arranging large annotated dataset is quite impossible. Moreover, the image quality is also inferior. The solution to this problem is transfer learning (TL), which helps to create an accurate model by starting the learning from patterns that have been already learned on solving different problems instead of learning from the scratch~\cite{raghu2019transfusion, pan2009survey}. Therefore, TL is an approach in DL and ML techniques, that allow us to transfer knowledge from one model to another. There are two steps in a TL process. The first step is to select a pre-trained model that is trained on a large scale of benchmark dataset, which is related to the problem we intend to solve. For instance, Keras offers a wide range of pre-trained network such as VGG, Inception, Xception, ResNet in the literature. The second step is to fine-tune the model considering the size and similarity of our dataset with the pre-trained model. For instance, if we have a considerable amount of dataset, which is different from the pre-trained model dataset. Therefore, it is wise to train the entire model. Nevertheless, for a small amount of dataset, we need to freeze most of the layers and train only a few layers.

In this study, we have utilized VGG series, XceptionNet and ResNet50 network in the TL process, where the weights are pretrained on ImageNet dataset. ImageNet consists of $1.2$ million training, $50,000$ validation and $100,000$ testing images and belonging to $1000$ classes. As it is observed from our workflow diagram in Fig.~\ref{workflow}-(c), the earlier layers of every CNN model is frozen, which is responsible for capturing more generic features. Then, we have retrained the latter layers of the network as fine-tuning by training on cervical cancer cells dataset to capture more dataset-specific features. Finally, we have fine-tuned our own fully connected classifier. Fig.~\ref{transfer} presents VGG16 network as an example, where the first few convolutional blocks use transferred parameters ($w_1, w_2, w_3 ... ,w_k$) from another VGG16 network that is trained on ImageNet dataset. 

For all the four CNN’s, the input size is ($224 \times 224 \times 3$), the learning rate is $10^{-3}$  for 50 epochs and then continued training for another 50 periods with learning rate $10^{-5}$, the batch size is 32 for the training set, batch size is one is for the testing set, and Adam optimizer is employed. Fig.~\ref{transfer}-(a) exhibits the whole TL process as an example on the VGG network, where the first few layers are pre-trained on ImageNet dataset, and latter convolutional blocks along with FCL are fine-tuned. Fig.~\ref{transfer}-(b) shows some representative feature maps extracted from various convolutional blocks of the VGG-16 network, which demonstrates the capability of TL process for extracting meaningful information from the images.

\begin{figure}[ht!]
\centering
\includegraphics[scale=.8]{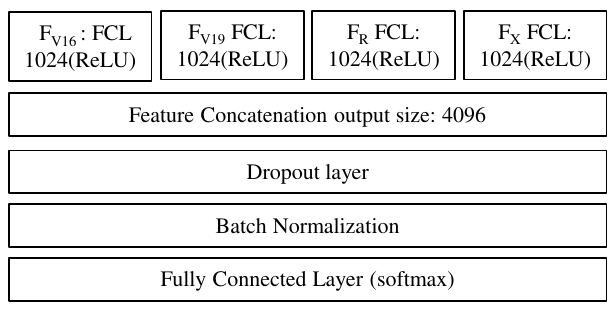}
\caption{Framework of the proposed hybrid feature fusion network.}
\label{ffn}
\end{figure}

\subsubsection{Late fusion technique}
Late fusion (LF) is one type of ensemble classifiers that relies on the maximum number of classifier decisions and then weights that decision to improve the classification performance. In this experiment, the classification result of four different DL models, namely, VGG16, VGG19, ResNet50, and XceptionNet, are combined using a majority voting technique, where each class is determined based on the highest number of votes received on that class. If $m = 1, 2, 3, …., X$ and $n= 1, 2, 3, ……, Y$, where $X$ is the number of classifiers, and $Y$ is the number of classes, the $i^{th}$ classifier's decision can be represented as $E(m,n)\in(0,1)$. The LF technique for majority voting can be described as follows,

\begin{equation}
\sum_{m=1}^{X} E(m, n) = max_{n=1}^Y  \sum_{m=1}^{X} E(m, n)
\end{equation}

\subsubsection{Feature fusion network}
Feature representation plays a vital role in image classification. We have observed that feature fusion (FF) is an efficient approach for cervical cytopathology cell image analysis. FF strategy combines multiple relevant features into a single feature vector, which contain rich information and contributes more descriptions than the initial input feature vectors. The traditional strategies for FF are serial and parallel FF~\cite{yang2003feature}. In a serial FF method, two features are concatenated into a single feature. For instance, two features $F_1$ and $F_2$ are extracted from an image with x, y vector dimension, then, fused feature is $F_s = (x+y)$. Whereas, parallel FF merges two components into a complex vector, $F_p= F_1+iF_2$ with $i$ indicating an imaginary component.

The problem with the above mentioned FF techniques is that they are unable to use original input features since they are creating new features. Moreover, they suffer from integrating multiple features. In our study, we have proposed an HDFF technique by integrating feature vectors from multiple CNN architectures. Fig.~\ref{ffn} shows our proposed DFF network, where $F_V16$, $F_V19$, $F_R$, $F_X$ are the normalized feature vectors extracted from the dense layer (FCL) with 1024 neurons of VGG16, VGG19, ResNet50 and XceptionNet. The FFN consists of one concatenation layer and one FCL layer with softmax activation function to integrate different features. Moreover, dropout and batch normalization layers are introduced to prevent overfitting and optimize training performance. The concatenation layer generates a vector of 4096 dimensions. If we consider $\bigcup$ for the concatenation operation, $F^n(i)$ indicates the $n$th feature vector. Then, the output vector of $i$th sample $F(i)$ can be written as
\begin{equation}
    F(i) = \bigcup_{i=1}^{4} F^n(i)
\end{equation}

\section{Experiments and Analysis}\label{sec:experiment}
\subsection{Dataset description}
\begin{figure*}[t!]
\centering
\includegraphics[scale=.88]{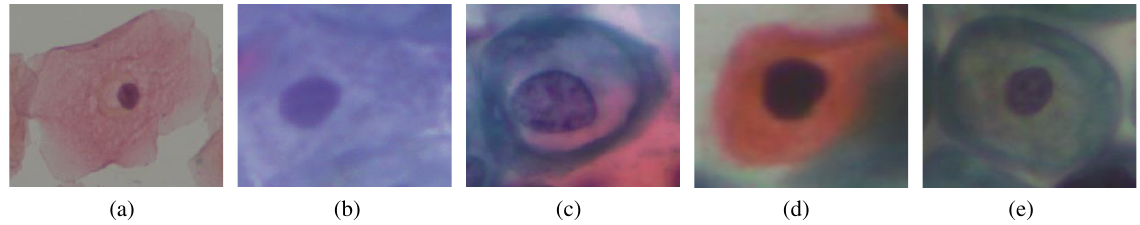}
\caption{An example of SIPAKMED database in five categories: (a) Superficial-Intermediate, (b) Parabasal, (c) Koilocytotic, (d) Dyskeratotic, (e) Metaplastic.}
\label{visualization}
\end{figure*}

To investigate the performance of our proposed DeepCervix network, we have applied publicly available SIPAKMED dataset consisting of 4049 annotated cervical pap smear cell images~\cite{plissiti2018sipakmed}. A set of dataset is displayed in Fig.~\ref{visualization}. Based on the cell appearance and morphology, expert cytopathologists classified the cells into five categories, such as superficial-intermediate, parabasal, koilocytotic, metaplastic and dyskeratotic. More precisely, Superficial-intermediate and parabasal cells can be further categorized as normal cells, koilocytotic and dyskeratotic cells are recognized as abnormal cells, and metaplastic cells are counted under benign cells. Table~\ref{num} provides the distribution of cells according to their classes.
\begin{table}[h!]
\small
\centering
\renewcommand{\arraystretch}{1.3}
\caption{Distribution of the SIPAKMED database}
\begin{tabular}{|l|l|c|}
\hline
\multicolumn{2}{|c|}{Category} & \begin{tabular}[c]{@{}c@{}}Number of Cells\end{tabular} \\ \hline
Superficial  & \multirow{2}{*}{Normal}   & 831  \\ \cline{1-1} \cline{3-3} 
Parabasal    &                           & 787  \\ \hline
Koilocytotic & \multirow{2}{*}{Abnormal} & 825  \\ \cline{1-1} \cline{3-3} 
Dyskeratotic &                           & 813  \\ \hline
Metaplastic  & Benign                    & 793  \\ \hline
\multicolumn{2}{|c|}{Total}              & 4049 \\ \hline
\end{tabular}
\label{num}
\end{table}

\subsection{Data setting}
SIPAKMED dataset comprises 4049 annotated cervical cell images. Among them, 60\% of the dataset in each class is used for training, 20\% is for validation, and 20\% is for testing. We have performed 5-class (superficial, parabasal, koilocytotic, metaplastic and dyskeratotic), 3-class (Normal, abnormal and benign) and 2-class (Normal and abnormal) classification of the dataset. Moreover, data augmentation techniques are used on the training set, which increases the training dataset by a factor of 6. The resulted training, validation and test dataset is shown in Table~\ref{datadis}.
\begin{table}[h!]
\small
\centering
\renewcommand{\arraystretch}{1.5}
\caption{The experimental data setting of SIPAKMED dataset}
\begin{tabular}{|l|c|c|c|}
\hline
\multicolumn{1}{|c|}{\multirow{2}{*}{Dataset}} & \multicolumn{3}{c|}{Total Number of Images} \\ \cline{2-4} 
\multicolumn{1}{|c|}{} & 5-Class & 3-Class & 2-Class \\ \hline
Training               & 16982   & 16989   & 13664   \\ \hline
Validation             & 811     & 811     & 652     \\ \hline
Test                   & 812     & 811     & 652     \\ \hline
\end{tabular}
\label{datadis}
\end{table}

\subsection{Experimental setup}
In this experiment, we have used Google Colaboratory, which is a cloud service based on Jupyter notebook, to train and test our model~\cite{bisong2019google}. Python 2 and 3 are pre-configured with many other ML libraries, such as Tensorflow, MatplotLib, Keras, PyTorch and OpenCV in Jupyter notebook. It provides run time with fully functional GPU (NVIDIA Tesla K80) in Colab environment to exercise DL. Moreover, the codes are protected in Google drive.

\subsection{Evaluation method}
To overcome the bias among the different algorithms, selecting a suitable evaluation metric is vital. Precision, recall, F1 score and accuracy are the most standard measures to evaluate the classification performance~\cite{sukumar2016computer}. The number of correctly identified samples among the all recognized representations are known as precision, whereas recall defines the ability of a classification model to recognize all the relevant samples. The F1 score combines both metrics, precision and recall, using the harmonic mean. Accuracy is the proportion of correctly predicted samples from the total number of samples. The mathematical expressions of the evaluation metrics are shown in Table~\ref{evaluationp} . In Table~\ref{evaluationp}, true positive (TP) is the number of accurately labeled positive samples, true negative (TN) is the number of correctly classified negative samples, the number of negative samples classified as positive are False positive (FP), and the number of positive instances predicted as negative is a false negative (FN).

\begin{table}[h!]
\small
\centering
  \renewcommand{\arraystretch}{1.2}
    \caption{Evaluation metrics}
    \begin{tabular}{l   |  c} 
      \hline     
      \textbf{Assessments} & \textbf{Formula}\\
      \hline
      Precision, $P$ & $\frac{TP}{TP + FP}$\\\\

      Recall, $R$ & $\frac{TP}{TP + FN}$\\\\

      F1 score &    $2 \times\frac{P\times R} {P + R}$\\\\

      Accuracy &      $\frac{TP + TN}{TP + TN + FP + FN}$\\
      \hline
      
    \end{tabular}
    \label{evaluationp}
\end{table}

\subsection{Results and analysis}

\subsubsection{Evaluation results}
To exam the performance of our proposed HDFF method, we have calculated the precision, recall, F1 score and accuracy of each individual fine-tuned DL models (VGG16, VGG19, ResNet-50, XceptionNet) along with late fusion (LF), where we have implemented the majority voting of diverse classifier (MVDC) and HDFF methods.  The performance results for the classification of cervical cells on the unseen test dataset are shown in Table~\ref{resultq}. The results are analyzed for binary class, 3-class and 5-class classification problems.\\

\begin{table}[h!]
\small
\centering
\renewcommand{\arraystretch}{1.5}
\caption{Performance analysis of the proposed HDFF method along with the base models. (Average Precision (Avg. P), Average Recall (Avg. R), Average F1 score (Avg. F1), Late Fusion (LF)}
\begin{tabular}{|c|l|l|l|l|l|}
\hline
\begin{tabular}[c]{@{}c@{}}Cl. Pro.\end{tabular} &
  CNN Models &
  Avg. P &
  Avg. R &
  Avg. F1 &
  \multicolumn{1}{c|}{\begin{tabular}[c]{@{}c@{}}Acc.\\ (\%)\end{tabular}} \\ \hline
\multirow{6}{*}{2-Class} & VGG16       & 1.00  & 1.00  & 0.998 & \textbf{99.85} \\ \cline{2-6} 
                         & VGG19       & 0.985 & 0.985 & 0.990 & 98.77 \\ \cline{2-6} 
                         & ResNet-50   & 0.995 & 0.995 & 0.990 & 99.38 \\ \cline{2-6} 
                         & XceotionNet & 0.980 & 0.980 & 0.980 & 98.31 \\ \cline{2-6}
                         & LF & 1.00 & 1.00 & 0.998 & \textbf{99.85} \\ \cline{2-6}
                         & HDFF        & 1.00  & 1.00  & 0.998 & \textbf{99.85} \\ \hline
\multirow{6}{*}{3-Class} & VGG16       & 0.976 & 0.970 & 0.973 & 97.90 \\ \cline{2-6} 
                         & VGG19       & 0.963 & 0.943 & 0.953 & 96.18 \\ \cline{2-6} 
                         & ResNet-50   & 0.963 & 0.950 & 0.956 & 96.18 \\ \cline{2-6} 
                         & XceptionNet & 0.923 & 0.963 & 0.880 & 89.64 \\ \cline{2-6}
                         & LF  & 0.987 & 0.980 & 0.980 & 98.52 \\ \cline{2-6} 
                         & HDFF        & 0.993 & 0.990 & 0.993 & \textbf{99.38} \\ \hline
\multirow{6}{*}{5-Class} & VGG16       & 0.983 & 0.981 & 0.980 & 98.27 \\ \cline{2-6} 
                         & VGG19       & 0.966 & 0.962 & 0.964 & 96.43 \\ \cline{2-6} 
                         & ResNet-50   & 0.964 & 0.958 & 0.960 & 96.06 \\ \cline{2-6}
                         & XceptionNet & 0.751 & 0.650 & 0.639 & 65.77 \\ \cline{2-6} 
                         & LF  & 0.986 & 0.986 & 0.986 & 98.64 \\ \cline{2-6}
                         & HDFF        & 0.992 & 0.990 & 0.990 & \textbf{99.14} \\ \hline
\end{tabular}
\label{resultq}
\end{table}

\begin{figure*}[t!]
\centering
\includegraphics[scale=.80]{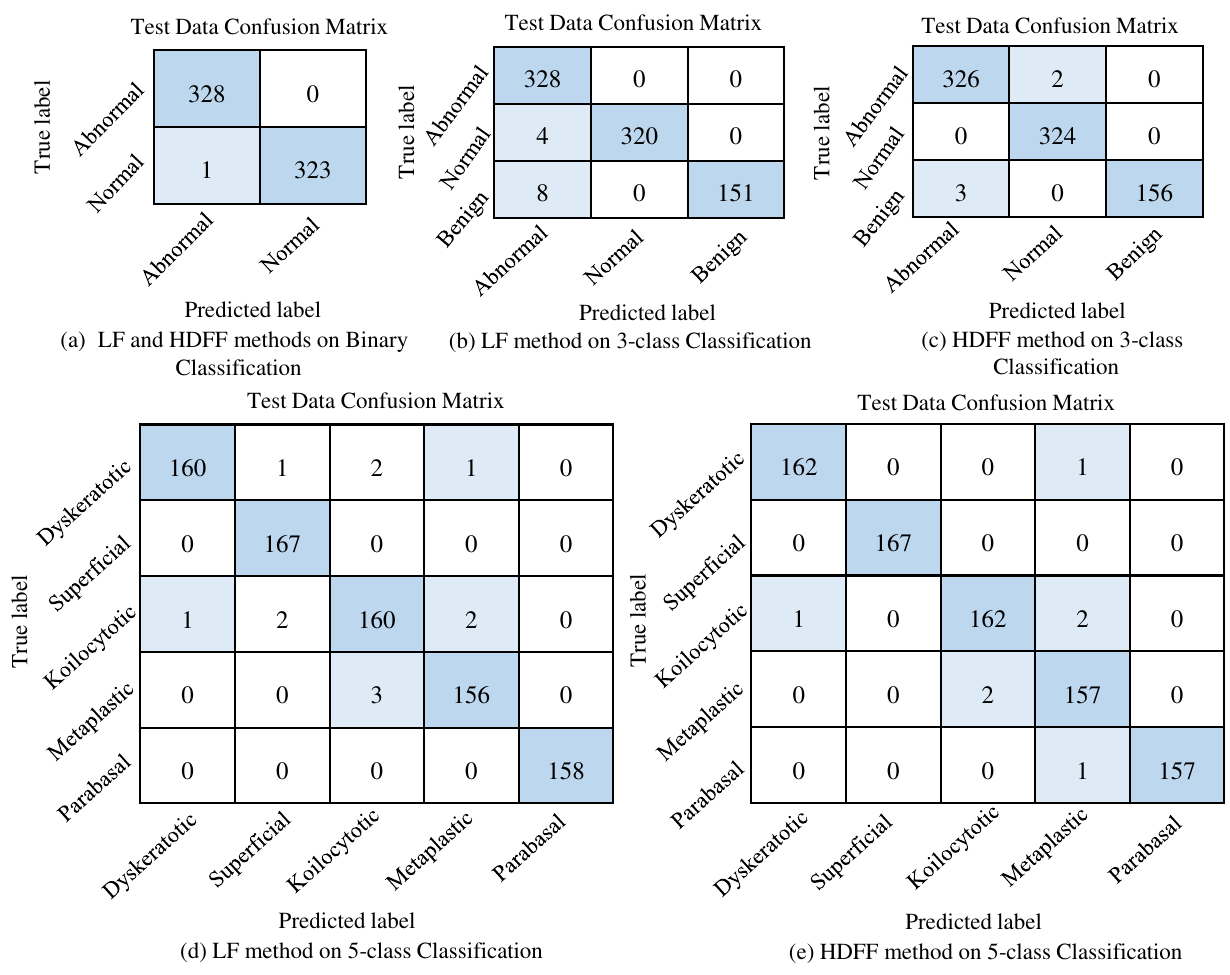}
\caption{The confusion matrix of the LF and HDFF methods for 2-class, 3-class and 5-class classification problem. }
\label{matrix}
\end{figure*}

\begin{figure*}[]
\centering
\includegraphics[scale=.80]{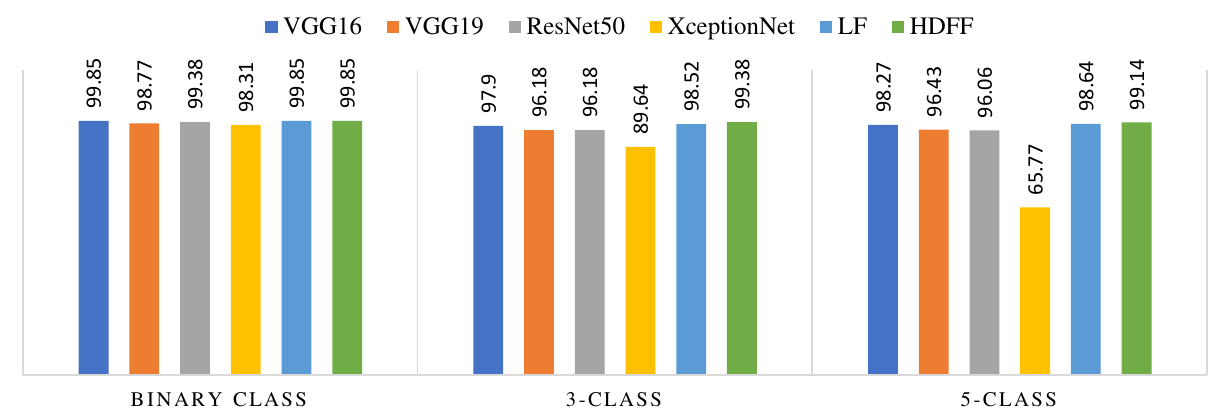}
\caption{Performance comparison among different TL models with HDFF and LF methods.}
\label{bar}
\end{figure*}

\emph{Binary classification:} In this case, we have classified the cervical cells into Normal and Abnormal (Table~\ref{num}). It is seen from Table~\ref{resultq} that, among the four DL models, VGG16 gives the highest average precision, recall, F1 score of 1.00, 1.00, 0.998, respectively, with an overall accuracy of 99.85\%. After VGG16, ResNet-50 gives the  classification accuracy of 99.38\%, with an average precision, recall and F1 score of 0.995, 0.995 and 0.990. Whereas, XceptionNet performs the least among them with an overall accuracy of 98.31\%. Moreover, MVDC based LF and HDFF techniques achieve a similar result as VGG16.\\

\emph{3-Class classification:} For ternary classification, we have classified the cervical cells into Normal, Abnormal and Benign class (Table~\ref{num}). It can be seen from Table~\ref{resultq} that VGG16 obtains the classification accuracy of 97.90\% with an average precision, recall and F1 score of 97.60\%, 97\% and 97.3\%. VGG19 and ResNet-50 provide the same average precision value of 0.963, recall value of 0.943, 0.950 and F1 score of 0.953, 0.956, respectively. Both of them obtain an accuracy of 96.18\%. However, XceptionNet shows the worst performance and contribute with an accuracy of 89.64\%. Additionally, LF technique obtains an accuracy of 98.52\%, with precision, recall and F1 value of 0.987, 0.980 and 0.980, respectively. Our advanced HDFF method obtains the highest classification accuracy of 99.38\% with an average precision, recall and F1 score of 0.993, 0.990, 0.993, respectively. \\

\emph{5-class classification:} In this experiment, we have analyzed the cervical cells into five classes (Table~\ref{num}). It is shown from Table~\ref{resultq} that the highest overall accuracy, precision, recall and F1 score is 99.14\%, 0.992, 0.990 and 0.990, obtained by HDFF technique, followed by LF method, VGG16, VGG19, ResNet50 and XceptionNet with an overall accuracy of 98.64\%, 98.27\%, 96.43\%, 96.06\% and 65.77\%, respectively. XceptionNet gives the worst performance with an average precision, recall and F1 score of 0.751, 0.650, 0.639, respectively.\\

The performance results in Table~\ref{resultq} illustrate that our proposed HDFF method (DeepCervix) obtains the highest classification accuracy for binary class, 3-class and 5-class classification problem. After the HDFF method, LF achieves the top classification results. Among the four DL models, VGG16 always provides superior performance, whereas the performance of XceptionNet degrades with the extension of number of classes. It is also observed that binary classification achieved the highest classification accuracy, followed by 3-class and 5-class classification problem.

\subsubsection{Visualized analysis}
To better illustrate the classification performance, we present confusion matrices of our proposed HDFF and LF methods in Fig.~\ref{matrix}. Moreover, Fig.~\ref{bar} shows the accuracy of each DL, LF, and HDFF models in histogram charts.
 
If we look at the confusion matrix for binary classification in Fig~\ref{matrix}-(a), it is seen that both of the models (HDFF and LF) can accurately recognize 328 images as abnormal and 323 images as normal, though one regular image is labeled as abnormal. According to Table~\ref{resultq}, both of the models obtained the same accuracy. For 3-class and 5-class classifications, the HDFF method has better recognition ability than the LF method. From Fig.~\ref{matrix}-(c) it is observed that the HDFF method can accurately recognize 326 images as abnormal, 324 images as normal, and 156 images as benign, whereas only five images are misclassified. For 5-class classification, the HDFF method accurately classified 805 images out of 812 images (Fig.~\ref{matrix}-(e)).

According to the histogram diagram in Fig.~\ref{bar}, it is recognized that all of the models obtained considerably very high accuracy for binary classification problems. As the number of classes increases, the overall accuracy for individual DL models decreases, whereas our proposed HDFF method shows good performance. For 3-class classification problem, the accuracy for the HDFF method is 99.38\%, which is 1.48\%, 3.2\%, 3.2\%, 9.74\%, 0.86\% higher than VGG16, VGG19, ResNet-50, XceptionNet and LF method, respectively. For 5-class classification, the highest classification accuracy is 99.14\%, achieved using HDFF method, which is an improvement of 0.87\% than VGG16, 0.5\% than LF, 2.71\% than VGG19, 3.08\% than ResNet50, and 33.37\% than XceptionNet.

\subsubsection{Performance comparison between HDFF method with existing researches using SIPAKMED dataset}
Table~\ref{Comparison} presents a comparative analysis of our proposed HDFF method with existing works to classify cervical cells using the SIPAKMED dataset. It is recognized from the table that our proposed HDFF method obtained the highest classification accuracies on binary and multiclass classification problems. For binary and 5-class classification problems, our method obtained 1.60\% and 0.19\% higher accuracies than the current studies. It is noticed that the 3-class classification problem has not been addressed in existing researches. 

\begin{table}[h!]
\small
\centering
\renewcommand{\arraystretch}{1.3}
\caption{Comparison of classification accuracies on SIPAKMED dataset}
\begin{tabular}{|c|l|c|c|}
\hline
Ref.   & \multicolumn{1}{c|}{Method}                                           & Class   & Accuracy \\ \hline
\cite{plissiti2018sipakmed} & CNN                                                                   & 5-Class & 95.35\%  \\ \hline
\cite{Shi2019GraphCN}   & \begin{tabular}[c]{@{}l@{}}Graph convolutional\\ network\end{tabular} & 5-Class & 98.37\%  \\ \hline
\cite{talo2019diagnostic}   & DenseNet-161                                                          & 5-Class & 98.96\%  \\ \hline
\cite{win2020computer} &
  \begin{tabular}[c]{@{}l@{}}Bagging Ensemble\\ Classifier\end{tabular} &
  \begin{tabular}[c]{@{}c@{}}2-Class\\ 5-Class\end{tabular} &
  \begin{tabular}[c]{@{}c@{}}98.25\%\\ 94.09\%\end{tabular} \\ \hline
Our method &
  HDFF &
  \begin{tabular}[c]{@{}c@{}}2-Class\\ 3-Class\\ 5-Class\end{tabular} &
  \begin{tabular}[c]{@{}c@{}}\textbf{99.85\%}\\ \textbf{99.38\%}\\ \textbf{99.15\%}\end{tabular} \\ \hline
\end{tabular}
\label{Comparison}
\end{table}

\subsubsection{Computational time}
In our experiment, first, we have trained the individual DL models (VGG16, VGG19, ResNet50, XceptionNet) and saving them with their weights separately. Then, we use those saved models and their weights and perform further training in the HDFF method stage. To train each DL model, it takes around six hours for 100 epochs (using google colab). To train the HDFF model by using the saved models requires only a few minutes(3 seconds per epoch). Though it requires quite a long time for training, the testing time is around 2.5 seconds for each cervical cells.

\subsection{Additional Experiment}
\subsubsection{Dataset}
Publicly available pap smear benchmark dataset (Herlev dataset)~\cite{jantzen2005pap}, consists of 917 single-cell images, is employed to evaluate our proposed HDFF method.  This dataset is divided into seven classes. These seven classes can be further classified into benign and malignant. The benign class consists of 242 images, and the malignant class consists of 675 images. The details of the dataset are given in Table~\ref{Herlev}.

\begin{table}[h!]
\small
\centering
\renewcommand{\arraystretch}{1.3}
\caption{Distribution of the Herlev dataset}
\begin{tabular}{|l|c|c|}
\hline
\multicolumn{2}{|c|}{Category}                    & \multicolumn{1}{l|}{Number of Cells} \\ \hline
Normal squamous       & \multirow{3}{*}{Normal}   & 74                                   \\ \cline{1-1} \cline{3-3} 
Intermediate squamous &                           & 70                                   \\ \cline{1-1} \cline{3-3} 
Columnar              &                           & 98                                   \\ \hline
Mild dysplasia        & \multirow{4}{*}{Abnormal} & 182                                  \\ \cline{1-1} \cline{3-3} 
Moderate dysplasia    &                           & 146                                  \\ \cline{1-1} \cline{3-3} 
Severe dysplasia      &                           & 197                                  \\ \cline{1-1} \cline{3-3} 
Carcinoma in situ     &                           & 150                                  \\ \hline
\multicolumn{2}{|c|}{Total}                       & 917                                  \\ \hline
\end{tabular}
\label{Herlev}
\end{table}

Our experiment took 60\% images of each class for training, 20\% is for validation, and the rest is for testing. Besides, the data augmentation technique is addressed on the training set, which increases the training dataset by a factor of 14. The resulting training, validation, and test dataset for 7-class and 2-class classification problems are given in Table~\ref{herlevdistribution}.
\begin{table}[h!]
\small
\centering
\renewcommand{\arraystretch}{1.3}
\caption{The experimental data setting of Herlev dataset}
\begin{tabular}{|l|c|c|}
\hline
\multicolumn{1}{|c|}{\multirow{2}{*}{Dataset}} & \multicolumn{2}{c|}{Total Number of Images} \\ \cline{2-3} 
\multicolumn{1}{|c|}{} & 7-Class & 2-Class \\ \hline
Training               & 8190    & 8235    \\ \hline
Validation             & 185     & 184     \\ \hline
Test                   & 186     & 184     \\ \hline
\end{tabular}
\label{herlevdistribution}
\end{table}

\subsubsection{Experimental Results on the Herlev dataset}
Table~\ref{resultherlev} presents the classification performance of four different DL models with the LF and HDFF methods. The four CNN models are accepted as a backbone network of LF and HDFF models. For binary classification of the Herlev dataset, it is observed that ResNet-50 provides the highest precision, recall, and F1 score for distinguishing the normal cervical cells from the abnormal one amid of the four CNN models, followed by VGG19, VGG16, and XceptionNet. Among the LF and HDFF methods, the HDFF method achieves the highest classification accuracy of 98.91\%, which is 1.08\% higher than the LF method.
 
For the 7-class classification of the Herlev dataset, ResNet-50 provides the highest classification accuracy of 83.87\% among the four CNN models, whereas XceptionNet performs the worst and gives an accuracy of 39.78\%. The LF approach reaches 86.02\% accuracy, with an average precision, recall, and F1 score of 0.887, 0.872, 0877, respectively. Moreover, our proposed HDFF method obtains the highest classification accuracy of 90.32\%, with an average precision, recall, and F1 score of 0.915, 0.911, and 0.916, respectively.
 
It is recognized that, for both the binary and multiclass classification problems, ResNet-50 obtains the highest classification accuracy among the four DL models. After ResNet50, the LF model achieves better results than the individual DL models, whereas the HDFF method obtains the highest classification accuracy.
\begin{table}[h!]
\small
\centering
\renewcommand{\arraystretch}{1.3}
\caption{Performance analysis of the proposed HDFF method along with the base models on Herlev dataset. (Average Precision (Avg. P), Average Recall (Avg. R), Average F1 score (Avg. F1), Late Fusion (LF)}
\begin{tabular}{|l|l|llll|}
\hline
Cl. Pro. &
  CNN Models &
  \multicolumn{1}{l|}{Avg. P} &
  \multicolumn{1}{l|}{Avg. R} &
  \multicolumn{1}{l|}{Avg. F1} &
  \multicolumn{1}{c|}{\begin{tabular}[c]{@{}c@{}}Acc.\\ (\%)\end{tabular}} \\ \hline
\multirow{6}{*}{2-Class} & VGG16       & 0.880 & 0.895 & 0.885 & 90.76 \\
                         & VGG19       & 0.910 & 0.845 & 0.870 & 90.76 \\
                         & ResNet-50   & 0.950 & 0.930 & 0.940 & 95.11 \\
                         & XceptionNet & 0.850 & 0.815 & 0.835 & 87.50 \\
                         & LF          & 0.985 & 0.960 & 0.975 & 97.83 \\
                         & HDFF        & 0.995 & 0.980 & 0.985 & \textbf{98.91} \\ \hline
\multirow{6}{*}{7-Class} & VGG16       & 0.684 & 0.641 & 0.929 & 61.29 \\
                         & VGG19       & 0.660 & 0.645 & 0.644 & 59.68 \\
                         & ResNet50    & 0.860 & 0.850 & 0.853 & 83.87 \\
                         & XceptionNet & 0.412 & 0.425 & 0.380 & 39.78 \\
                         & LF          & 0.887 & 0.872 & 0.877 & 86.02 \\
                         & HDFF        & 0.915 & 0.911 & 0.916 & \textbf{90.32} \\ \hline
\end{tabular}
\label{resultherlev}
\end{table}

\subsubsection{Performance comparison between HDFF method with existing researches using Herlev dataset}
Table~\ref{Comparisonherlev} compares the performance results of existing studies with our proposed HDFF method in terms of overall classification accuracy for 2-class and 7-class classification problems. A higher accuracy value indicates a higher rate of correct classifications. It is observed from the table that most of the existing work perform binary class classification tasks, and they obtain accuracy above 90\%. However, only a few papers addressed both the binary and multiclass classification of the Herlev dataset. For the multiclass classification problem, the classification accuracy is between 68.54\% to 95.9\%. \cite{allehaibi2019segmentation} obtains the highest accuracy for 7-class classification, but it requires pre-segmented cervical cell images. It is further observed from Table~\ref{Comparisonherlev} that our proposed HDFF method outperforms existing methods in most cases, which shows the robustness of our proposed algorithm.
\begin{table}[h!]
\footnotesize
\centering
\renewcommand{\arraystretch}{1.3}
\caption{Comparison of classification accuracies on Herlev dataset ( BPNN (Back propagation neural network), LSSVM (Least-squares support-vector machines), HVCA (Hybrid variational convolutional autoencoder), ETL (Ensembled transfer learning), Cl. Pro.(Classification problem), Acc (Accuracy))}
\begin{tabular}{|l|l|c|c|}
\hline
\multicolumn{1}{|c|}{Ref.} & \multicolumn{1}{c|}{Method}                                                                      & Cl. Pro.                                                  & Acc                                                       \\ \hline
\cite{singh2015neural}                          & BPNN                                                                                             & 3-Class                                                   & 79\%                                                      \\ \hline
\cite{sarwar2015hybrid}                       & Hybrid ensemble                                                                                  & \begin{tabular}[c]{@{}c@{}}2-Class\\ 7-Class\end{tabular} & \begin{tabular}[c]{@{}c@{}}98\%\\ 78\%\end{tabular}       \\ \hline
\cite{bora2016pap}                         & AlexNet \& LSSVM                                                                                 & 2-Class                                                   & 94.61\%                                                   \\ \hline
\cite{taha2017classification}                          & AlexNet \& SVM                                                                                   & 2-Class                                                   & 99.19\%                                                   \\ \hline
\cite{wieslander2017deep}                         & VGG16 \& ResNet                                                                                  & 2-Class                                                   & 86\%                                                      \\ \hline
\cite{zhang2017deeppap}                          & CNN \& TL                                                                                        & 2-Class                                                   & 98.3\%                                                    \\ \hline
\cite{nanni2017handcrafted}                          & CNN \& TL                                                                                        & 2-Class                                                   & 95.1\%                                                    \\ \hline
\cite{gautam2018considerations}                          & AlexNet \& TL \& DT                                                                              & \begin{tabular}[c]{@{}c@{}}2-Class\\ 7-Class\end{tabular} & \begin{tabular}[c]{@{}c@{}}99.3\%\\ 93.2\%\end{tabular}   \\ \hline
\cite{ lin2019fine}                          & Morphology \& CNN                                                                             & \begin{tabular}[c]{@{}c@{}}2-Class\\ 7-Class\end{tabular} & \begin{tabular}[c]{@{}c@{}}94.5\%\\ 64.5\%\end{tabular}   \\ \hline
\cite{allehaibi2019segmentation}                          & \begin{tabular}[c]{@{}l@{}}VGG-like network\\ (Segmened image)\end{tabular}                      & \begin{tabular}[c]{@{}c@{}}2-Class\\ 7-Class\end{tabular} & \begin{tabular}[c]{@{}c@{}}98.10\%\\ \textbf{95.9\%}\end{tabular}  \\ \hline
\cite{promworn2019comparisons}                          & \begin{tabular}[c]{@{}l@{}}DenseNet161\end{tabular} & \begin{tabular}[c]{@{}c@{}}2-Class\\ 7-Class\end{tabular} & \begin{tabular}[c]{@{}c@{}}94.38\%\\ 68.54\%\end{tabular} \\ \hline
\cite{khamparia2020dcavn}                          & HVCA                                                                                             & 2-Class                                                   & \textbf{99.4\%}                                                    \\ \hline
\cite{khamparia2020internet}                          & Pretrained ResNet50                                                                              & 2-Class                                                   & 97.89\%                                                   \\ \hline
\cite{xue2020application}                          & ETL                                                                                              & 2-Class                                                   & 98.37\%                                                   \\ \hline
Our method                 & HDFF                                                                                             & \begin{tabular}[c]{@{}c@{}}2-Class\\ 7-Class\end{tabular} & \begin{tabular}[c]{@{}c@{}}98.91\%\\ 90.32\%\end{tabular}           \\ \hline
\end{tabular}
\label{Comparisonherlev}
\end{table}

\begin{figure*}[]
\centering
\includegraphics[scale=.85]{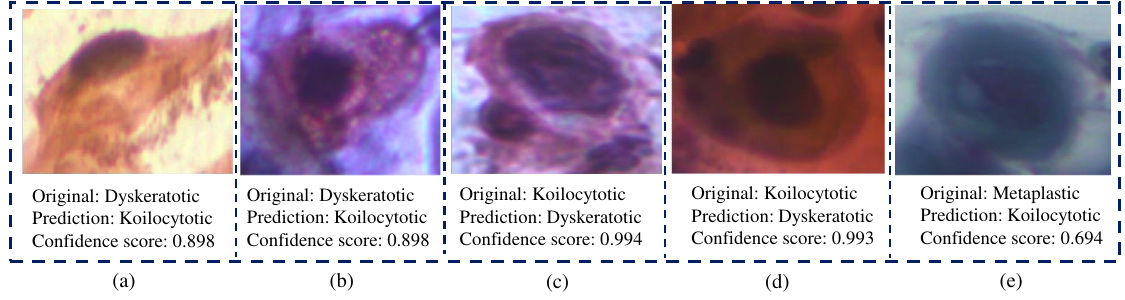}
\caption{Examples of misclassified cervical cells from SIPAKMED dataset.}
\label{miss}
\end{figure*}

\begin{figure*}[]
\centering
\includegraphics[scale=.85]{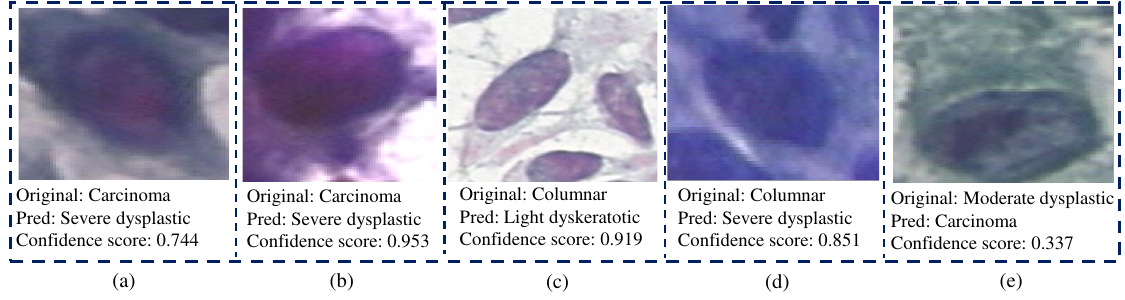}
\caption{Examples of misclassified cervical cells from Herlev dataset.}
\label{missherlev}
\end{figure*}

\section{Discussion}\label{discu}
Lately, the advancement of DL is solving critical tasks in the medical domain. Classification of cervical cells can help identify the cancerous subjects early, which is a significant step to prevent cervical cancers. This study proposes the HDFF method (DeepCervix) to classify the cervical cells on the SIPAKMED and Herlev datasets and obtained excellent results.

Imaging modality, image quality, dataset distribution, model structure, complexity, loss function, optimization function and number of epochs are some critical factors that influence a model's performance. When we observe the performance metrics for the SIPAKMED dataset in Table~\ref{resultq}, VGG16 performs relatively well compared to ResNet50, VGG19, and XceptionNet. Therefore, a shallow network performs better than a very deep network for the SIPAKMED dataset. If we consider the network architecture for VGG16, it contains very small receptive fields, which enables to have more weight layers and thus to improve performance. The LF model based on MVDC shows a slight improvement in the overall result, but it cannot always guarantee leading performance. Besides, the HDFF method can effectively improve the classification performance and provides the best result. It is observed from Fig.~\ref{matrix} that the HDFF method can correctly classify 805 images out of 812 images in a 5-class classification task. It is also observed that Koilocytotic and metaplastic are challenging cells to classify. For the Herlev dataset (Table~\ref{resultherlev}), unlike SIPAKMED, ResNet-50 performs better than other DL models. Therefore, it is observed that, for highly imbalanced and small datasets, ResNet-50 is preferable. Besides, the best performance is obtained by the HDFF method for 2-class and 7-class classification problems.

Fig.~\ref{miss} and Fig.~\ref{missherlev} provide examples of misclassified cervical cells on the SIPAKMED and Herlev dataset for the 5-class and 7-class classification problem. It can be seen from Fig~\ref{miss}-(a) that, for the dyskeratotic class image, the cell boundary and nucleus are hard to distinguish and are wrongly listed as Koilocytotic with a confidence score of 0.898. For Fig.~\ref{miss}-(b),(c) the Dyskeratotic and Koilocytotic class image looks identical with the invisible nucleus boundary and misclassified as koilocytotic and Dyskeratotic, respectively. Fig.~\ref{miss}-(d) reveals that the dark stained koilocytotic cell is misclassified as Dyskeratotic. From Fig.~\ref{miss}-(e), it can be found that the content of the Metaplastic cell is too dark to identify the cell and nucleus region and misclassified as koilocytotic with a confidence score of 0.694. According to Fig.~\ref{missherlev}-(a),(b) two dark-stained carcinoma images are labeled as severe dysplastic. In Fig.~\ref{missherlev}-(c),(d) two columnar images, which look very different to each other, are misclassified as light and severe dysplastic. It can be seen from Fig.~\ref{missherlev}-(e) that a moderate dysplastic cell image is misclassified as carcinoma. For all the misclassified images, it is recognized that none of them contain adequate information about a cell.

\section{Conclusion and Future work}\label{sec:conclu}
This study proposes a deep learning-based HDFF and LF method to classify cervical cells. It is observed from the performance metrics that the HDFF method achieves higher classification accuracies compared to the LF method. Unlike other methods that rely on pre segmentation of cytoplasm/nucleus and hand-crafted features, our proposed method offers end-to-end classification of cervical cells using deep features. SIPAKMED and Herlev datasets are utilized to evaluate the performance of our proposed model. For the SIPAKMED dataset, we have obtained the state-of-the-art accuracy of 99.85\%, 99.38\%, and 99.14\% for 2-class, 3-class, and 5-class classification problems. We have reached 98.91\% accuracy for the Herlev dataset for a binary classification problem and 90.32\% for the 7-Class classification problem. 

Though our method provides very good performance, there are a few limitations. First of all, despite the high accuracy of the SIPAKMED dataset, the performance of our method degrades for 7-class classification on the Herlev dataset. An ideal screening system should not miss any abnormal cells. To overcome this for the multiclass classification problem, we could have integrated pre-segmented cell features into our model. Secondly, for our HDFF method, we have investigated four DL models, fine-tuned them, and integrate their features to get the final model. In the future, we can investigate other DL models and compare their results for the multiclass classification accuracy. Thirdly, our proposed method should be generalized for the classification involving cell overlapping. Finally, poison noise is a critical factor for cervical cell images that degrades model performance. Therefore, the denoising methods, such as adaptive wiener filter~\cite{deepa2016study} in the preprocessing step can be implemented to improve the model's overall performance.

\bibliographystyle{unsrt}  

\bibliography{Mamun}
\section*{Acknowledgements}
This work is supported by the ``National Natural Science Foundation of China'' (No. 61806047), 
the ``Fundamental Research Funds for the Central Universities'' (No. N2019003) and
the ``China Scholarship Council'' (No. 2018GBJ001757). 
We also thank M.E. Dan Xue and B.E. Xiaomin Zhou in the previous work of this research. 
We thank Miss Zixian Li and Mr. Guoxian Li for their important support and discussion 
in this work.

\end{document}